\documentclass[12pt,oneside,final,a4paper]{article}

\title{Quantum statistics via perturbation effects of preparation procedures}

\author{Andrei Khrennikov\\
International Center for Mathematical Modeling\\ in Physics and Cognitive sciences \\
MSI, V\"axj\"o
University, S-35195, Sweden}

\begin{document}
\maketitle

\begin{abstract}
We study the following problem: Is it possible to explain
the quantum interference of probabilities in the purely 
corpuscular model for elementary particles? We demonstrate
that (by taking into account perturbation effects of measurement
and preparation procedures) we can obtain $\cos\theta$-perturbation
(interference term) in probabilistic rule connecting preparation procedures 
for purely corpuscular objects. On one hand, our investigation demonstrated
that there is nothing special in so called `quantum probabilities':
the right choice of statistical ensembles gives the possibility to escape
all `pathologies'. On the other hand, we found that the standard trigonometric
interference of alternatives (observed, in particular, in quantum mechanics)
is not the unique possibility to extend (disturb) the conventional probabilistic rule
for addition of alternatives. There exist two other  probabilistic rules that
connect three preparation procedures: hyperbolic and hyper-trigonometric interferences.
\end{abstract}

\section{Introduction}

It is well known that the quantum probabilistic rule
for interference of alternatives differs from the conventional
probabilistic rule. For `conventional probabilities', we have 
\begin{equation}
\label{CL}
{\bf P}= {\bf P}_1 + {\bf P}_2,
\end{equation}
for `quantum probabilities', we have:
\begin{equation}
\label{Q}
{\bf P}= {\bf P}_1 + {\bf P}_2+ 2\sqrt{{\bf P}_1 {\bf P}_2} \cos \theta.
\end{equation}
This difference in statistics is still one of the mysteries of 
modern physics. There are various explanations of the appearance 
of the $\cos \theta$-factor in quantum statistics. The common viewpoint 
is that new probabilistic behaviour which is observed in experiments
with elementary particles is a consequence of the wave feature of these
physical systems; see, for example,  [1]-[18]. It is also commonly assumed that
it would be impossible to explain
the appearance of the $\cos \theta$-factor by using the purely corpuscular model
for elementary particles. In fact, probabilistic transformation (\ref{Q}) was the main
reason to introduce the principle of complementarity, Bohr [2]. 

In this paper we demonstrate that, despite the common opinion, 
probabilistic rule (\ref{Q}) can be derived in the purely corpuscular
model as a consequence of the perturbation effects of preparation procedures.
In particular, by using our method it is possible to simulate the interference 
of alternatives for macro-systems (for example, a kind of two-slit experiment
for macro-balls). Our investigation was strongly motivated by the first chapters
of Dirac's book [1]. There P. Dirac investigated the roots of quantum behaviour.
He paid the large attention to the role of perturbations induced by preparation
and measurement procedures. He rightly pointed out that the magnitude of these
perturbations plays the crucial role in the transition from the classical theory to the
quantum theory.
However, he did not pay attention to the fact that these perturbations could produce  quantum 
probabilistic behaviour, (\ref{Q}), in the purely corpuscular model. Therefore
he, as many others, must use the wave particle duality (in particular, split of photons
in the two slit experiment) to explain the origin of $\cos \theta$-factor in the 
quantum probabilistic rule (\ref{Q}): `If the two components are now made to interfere,
we should require a photon in one component to be able to interfere with one in the other', [1].

We also pay the main attention to perturbation effects of preparation procedures.
We show that statistical deviations of sufficiently high magnitude
induce the $\cos\theta$-factor 
for probability distributions of purely corpuscular objects.\footnote{Statistical deviations
of negligibly small magnitude induce the conventional probabilistic rule.} An unexpected consequence
of our analysis of possible probabilistic transformations
(induced by preparation procedures) is that, beside trigonometric
interference of alternatives, (\ref{Q}), we may obtain 
hyperbolic interference:
\begin{equation}
\label{QA}
{\bf P}= {\bf P}_1 + {\bf P}_2 \pm 2\sqrt{{\bf P}_1 {\bf P}_2} \cosh \theta\;.
\end{equation}
This interference can also be simulated.

Of course, we do not claim that our investigation implies that elementary particles
are purely corpuscular objects. It may be that they have wave features and the principle
of complementarity represents the important aspect of quantum reality. However, we definitely
proved that quantum probabilistic rule (\ref{Q}) does not imply that we must use this principle
to describe elementary particles (compare with Dirac [1], Bohr [2], Schr\"odinger [4], [5], 
Feynman [7];
compare also with [13]-[18]).

Through this paper we use the realist model (see [9], [10], [13] for the details)
by that physical systems have
objective properties. Values of physical observables can be considered as properties
of the object. Of course, we do not claim that such a model can be used to describe
quantum particles. However, we demonstrated that nontrivial `quantum interference'
of alternatives can be obtained even in the realist model. 

We remark that (due to taking into account the perturbation
effects of preparation procedures) our realist objective model does not differ
essentially from so called contextualist model, see, for example,  [10], [13]. 
In the latter model values 
of physical observables are merely determined by the context of an experiment. 
Perturbation effects of measurement procedures that play the crucial role
in our investigation can be considered as the description of the experimental
arrangement (context of an experiment). In principle, we can easily rewrite 
this paper in the purely contextualist framework. Here we should use sub-quantum model
with hidden variables, see [13]. Instead of perturbation effects for physical
variables, we can consider perturbation effects for hidden variables. In such a way
we can reproduce all results of this paper. 

In the objective realist framework perturbation effects are merely associated with preparation procedures 
(state=ensemble preparations).  Preparation procedures perturb some (objective) properties of physical 
systems. By measurement we find numerical values describing these properties.
In the contextualist framework perturbation effects are merely 
associated with measurement procedures. We cannot consider physical observables as numerical
representations of properties of the object. The observed 
numerical values are created in the process of interaction between physical systems and measurement
devices. These values depend not only on physical systems, but on the whole experimental arrangement.
Nevertheless, such an interpretation of results of measurements does not exclude the possibility
to provide a sub-quantum deterministic description, hidden variables (HV) model that reproduces
predictions of quantum theory, see [13] for the details. The probabilistic model presented
in this paper can be considered as such a HV-model (contextualist HV-model).

As we have already remarked, our investigation
could not be considered as the crucial argument against the wave particle dualism.
The same can be said about the choice between the realist objective and contextualist
interpretations of quantum mechanics. We just demonstrated that `quantum 
probabilistic interference' could be obtained even in the realist objective model
for corpuscular systems.

In this paper we discuss perturbation effects of preparation and measurement procedures.
We remark that we do not follow to W. Heisenberg [6]; we do not study perturbation
effects for individual measurements. We discuss statistical (ensemble) deviations
induced by perturbations.\footnote{Such an approach implies the statistical viewpoint to
Heisenberg uncertainty relation: the statistical dispersion principle, 
see Ballentine [14], [15] for the details.}

We do not consider the relation
of this model to the EPR-Bell considerations, [19], [20]. There is the large diversity
of opinions on the origin of experimental violations of Bell's inequality, see, for example, 
[19]-[26].
Suppose that these violations really imply the impossibility to use objective realist
interpretation of results of observations. Then our analysis would imply that 
EPR-Bell considerations was really the new step in the development of quantum theory
with implications that, in fact, could not be obtained on the basis of the original formalism
(despite the common opinion). Really, practically all founders of quantum theory (for example,
De Broglie [27], Schr\"odinger [4], Dirac [1], Bohr [2], von Neumann [3], Bohm [28]) were sure that
interference of probabilistic alternatives (that was demonstrated in the two slit 
experiment) could not coexist with purely corpuscular model for elementary particles. Especially 
extreme viewpoint was presented in works of  Schr\"odinger. Finally, he criticized
even attempts to use `classical notion of a particle' in quantum theory [5], see
[29] for the detailed analysis \footnote{However, compare
with: `The compulsion to replace the simultaneous happenings as indicated directly by the theory,
by alternatives, of which the theory is supposed to indicated the respective probabilities, arises from
the conviction that {\bf what we really observe are particles - that actual events always 
concern particles not waves.}' see citation in [30], p.376,  of Shr\"odinger's notes
for a seminar he was giving in Dublin in 1952; here bold shrift is given by me.}

On the other hand,
De Broglie [27], Dirac [1], Bohm [2], Feynman [7] tried to combine wave and corpuscular 
features of elementary particles by using different models. De Broglie imagined
elementary particle as a singularity in the wave. It seems that in this model `the mother
wave produces a child-particle' by varying the space-time distribution. Bohm considered
the pilot wave and elementary particle as indivisible whole. Here the wave is not more 
the mother of a particle. Von Neumann, Dirac, Feynman and the majority of
quantum community considered corpuscular objects having unusual physical `properties.'
The wave features of elementary particles were exhibited via superposition of alternatives in 
quantum state. On the other hand, it seems that Einstein  was the adherent of the 
corpuscular model with the objective realist interpretation of physical observables, see [30],
[31] for the detailed analysis.

In the connection with the EPR-Bell considerations we remark that (despite the common opinion)
experimental violations of Bell's inequality can peacefully coexist with local realist HV-model
with the contextualist interpretation of physical observables, see [13] and [32]. 

Finally, we note that HV-models are typically not taken into account (especially
by `real physicists'), because it is not useful to construct a complicated HV-model
just to reproduce the standard results of (more simple) quantum formalism. In particular,
such an argument is one of the main motivations for the rejection the Bohmian mechanics. In the 
opposite to such `repetitive'  HV-models, our model not only give the possibility to
reproduce the standard quantum interference of probabilistic alternatives, but also 
 predict new (hyperbolic) interference.

\section{Statistical deviations produced by perturbations}

\subsection{Conventional probabilistic rule}
Let $\Omega$ be an ensemble of physical systems having two physical properties 
$a$ and $b.$ These properties supposed to be {\it{objective}} properties of a 
physical system. To simplify considerations, 
we suppose that these properties can be described by dichotomic variables
$a=0,1$ and $b=0,1.$ 

This situation can be described by the conventional probability model 
(Kolmogorov axiomatics, [33]). Here ($\Omega, {\cal F}, {\bf P}_\Omega$) is a probability space: 
${\cal F}$ is a $\sigma$-algebra of subsets of 
$\Omega$ and ${\bf P}_\Omega$ is a $\sigma$-additive probabilistic measure on 
${\cal F}; a, b :\Omega\rightarrow\{0,1\}$ are random variables. In our considerations
the right choice of ensembles of physical systems will play the crucial role. Therefore
we prefer to use for all probabilities indexes of corresponding ensembles.\footnote{In Kolmogorov's
axiomatics $\Omega$ is an arbitrary (abstract) set. However, in physical modeling it is important 
to use the right realization of this set. In our investigation $\Omega$ will be
an ensemble of physical systems (for example, elementary particles). We underline that 
typically in mathematical works $\Omega$ is realized as the space of all possible sequences of trials,
see [18] for the details.}

We set 

$\Omega^b_i=\{ \omega\in\Omega:b(\omega)=i\}, i=0,1.$ 

As usual, we define conditional probabilities (Bayes' formula):
$$
{\bf P}_{\Omega_i^b}(O)={\bf P}_\Omega(O/\Omega^b_i)=
\frac{{\bf P}_\Omega(O \bigcap \Omega^b_i)}{{\bf P}_\Omega(\Omega^b_i)}, i=0,1, \; O \in {\cal F} .
$$

We shall be interested in conditional probabilities: 
$$
p_{ij}^{a/b}={\bf P}_\Omega(a=j/b=i) .
$$
As usual we have:
$$
p_{ij}^{a/b}=p_{ij}^{ab}/p_i^b,
$$ 
where $p_{ij}^{ab}={\bf P}_\Omega(a=j, b=i)$ and $p_i^b={\bf P}_\Omega(\Omega_i^b).$

Additivity of ${\bf P}_\Omega$ 
and the definition of conditional probabilities imply the formula of total probability:
$$
{\bf P}_\Omega(a=j)= {\bf P}_\Omega(b=0) {{\bf P}}_{\Omega_0^b}(a=j) +{\bf P}_\Omega(b=1) {{\bf P}}_{\Omega_1^b}(a=j),
$$
or
\begin{equation}
\label{B}
p_j^a \equiv {\bf P}_\Omega(a=j) =p_0^b p_{0j}^{a/b} + p_1^b p_{1j}^{a/b}, j=1,2 .
\end{equation}
We remark that the formula of total probability can be used as the {\it prediction rule}:
if we know probabilities for the variable $b$ and conditional probabilities for
the variable $a,$ then we can predict probabilities for the $a.$ We call this rule,
the {\it conventional probabilistic rule.}

\subsection{Quantum probabilistic rule}
The quantum mechanical formalism does not reproduce the formula of total probability;
we cannot use the conventional probabilistic rule to predict probabilities in experiments
with elementary particles. The right hand side of (\ref{B}) is perturbed by a 
$\cos\theta$-factor. Often such a difference in transformation laws is interpreted as 
an evidence that objective realism could not be used in quantum framework.
It seems that there is something mysterious in the appearance 
$\cos\theta-$perturbation in `quantum formula of total probability'! 
Our aim is to provide probabilistic explanation of the appearance of this $\cos \theta$-factor.

Let $\varphi$ be a quantum state and let $\Omega$ be a statistical ensemble 
of quantum systems prepared for $\varphi.$ Let $a$ and $b$ be represented by 
operators $\hat{a}$ and $\hat{b}$ and $\{\psi_i\}_{i=0,1}$ and $\{\varphi_i\}_{i=0,1}$ be 
orthonormal bases of eigenvectors for $\hat{a}$ and $\hat{b}.$ 
We can expend the quantum state $\varphi$ with respect to basis 
$\{ \phi_i \}_{i=0,1}$
$$
\varphi = \sqrt{p_0^b} \varphi_0 + c^{i\xi} \sqrt{p_1^b}\varphi_1\;.
$$
As $(\varphi, \varphi)=1,$  we have $p_0^b+ p_1^b=1.$ By Born's probabilistic
interpretation, $p_i^b ={\bf P}_\varphi (b=i) \equiv {\bf P}_\Omega(b=i).$

We can also expend each $\varphi_i$ with respect to the basis $\{\psi_i\}_{i=0,1}:$
$$
\varphi_0 = e^{i\gamma_1}  \sqrt{q_{00}^{a/b}} \psi_0 + e^{i\gamma_2} \sqrt{q_{01}^{a/b}}
\psi_1,
$$
$$
\varphi_1 = e^{i\gamma_3} \sqrt{q_{10}^{a/b}}\psi_0+e^{i\gamma_4}
\sqrt{q_{11}^{a/b}}\psi_1,
$$ 
where $\xi, \gamma_1, \ldots, \gamma_4$ are phases. 
Here  (for values $j=0,1)$

$q_{ij}^{a/b}= {{\bf P}}_{\varphi_i}(a=j)$ (for states $i=0,1).$ 

We remark that, as 
each $\varphi_i$ is normalized, we have:
\begin{equation}
\label{PR}
q_{00}^{a/b} + q_{01}^{a/b}=1 , \; q_{10}^{a/b} + q_{11}^{a/b}=1 .
\end{equation}
This is the standard normalization conditions for probabilities of alternatives.
However, probabilities for quantum states (statistical ensembles used in quantum experiments)
satisfy to another normalization condition. Orthogonality of eigenvectors corresponding to physical
observables implies:
\begin{equation}
\label{PRD}
q_{00}^{a/b} + q_{10}^{a/b}=1 , \; q_{01}^{a/b} + q_{11}^{a/b}=1 .
\end{equation}
This is so called condition of double stochasticity, see [10] for the details.

The standard calculations in the linear space imply that the quantum 
probabilistic (prediction) rule has the form
\begin{equation}
\label{q0}
p_0^a = p_0^b q_{00}^{a/b} + p_1^b q_{10}^{a/b} + 2\; \sqrt{p_0^b p_1^b q_{00}^{a/b} q_{10}^{a/b}}\cos\theta\;,
\end{equation}
\begin{equation}
\label{q1}
p_1^a = p_0^b q_{01}^{a/b} + p_1^b q_{11}^{a/b} - 2\; \sqrt{p_0^b p_1^b q_{01}^{a/b} q_{11}^{a/b}} \cos\theta\;.
\end{equation}

It is cruicial for our further considerations that the probabilities $q_{ij}^{a/b}$ 
(corresponding to quantum states $\varphi_i)$  in general are not equal to 
probabilities $p_{ij}^{a/b}$ with respect to subensembles 
$\Omega_i^b, i=0,1,$ of the ensemble $\Omega$ for the quantum state $\varphi.$ 
In fact, $q_{ij}^{a/b}$ is the probability that $a=j$ in the ensemble $\tilde{\Omega}_i^b$ 
prepared for the quantum state $\varphi_i.$  Therefore there is nothing surprising that
conventional probabilistic rule (\ref{B}) differs from quantum probabilistic rule 
({\ref{q0}}), (\ref{q1}): 
probabilities $p_i^a, p_i^b$ with respect to the ensemble $\Omega$ need not be 
connected with the aid of probabilities $q_{ij}^{a/b}$ with respect 
to new ensembles $\tilde{\Omega}_i^b$ by the ordinary formula of total probability. 
We shall explain how the formula of total probability produces the quantum probabilistic rule
as a consequence of difference between the probabilities 
$p_{ij}^{a/b}$ and $q_{ij}^{a/b}.$

\subsection{Statistical deviations}
In general, we have three preparation procedures 
${\cal E}, {\cal E}_0, {\cal E}_1$ such that ${\cal E}$ produces an 
ensemble $\Omega$ of physical systems and ${\cal E}_0=F_0$ and ${\cal E}_1=F_1$ 
are filters with respect to some property $b(=0,1)$ that produce ensembles
$\tilde \Omega_ 0^b$ and $\tilde \Omega _1^b.$ The $p_i^a, p_i^b,$ are the probability distributions 
of $a$ and $b$ for the ensemble $\Omega; $ the $ q_{ij}^{a/b}$ are  probability distributions 
of  $a$ for the ensembles  $\Omega_i^b, i=0,1.$

Preparations of ensembles $\tilde{\Omega}_i^b, i=0, 1,$ can be realized with the aid of 
filters $F_i, i=0, 1,$ which select particles from $\Omega$ having the property 
$b=i, i=0, 1.$ These 
selections perturb physical systems. The original distribution of $a$ in $\Omega$ is 
changed. 

We do not restrict our considerations to quantum experiments. We consider arbitrary preparation
procedures for macro as well as micro-systems. We start with some evident 
manipulation with the ordinary formula of total probability 
for the ensemble $\Omega$:
$$
p_j^a = p_0^b  p_{0j}^{a/b}+ p_1^b p_{1j}^{a/b}= p_0^b q_{0j}^{a/b} + p_1^b q_{1j}^{a/b} + \delta_j(a,b),
$$
where
$$
\delta_j(a,b)=p_0^b(p_{0j}^{a/b}-q_{0j}^{a/b})+p_1^b(p_{1j}^{a/b}-q_{1j}^{a/b})
=2\sqrt{p_0^b p_1^b q_{0j}^{a/b}q_{1j}^{a/b}}\lambda_j,
$$
where
\begin{equation}
\label{L}
\lambda_j \equiv \lambda_j(a,b)=\frac{p_0^b(p_{0j}^{a/b}-q_{0j}^{a/b}) + p_1^b (p_{1j}^{a/b}-q_{1j}^{a/b})}{2\sqrt{p_0^b p_1^b q_{0j}^{a/b} q_{1j}^{a/b}}}
\end{equation}

We note that if $p_{0j}^{a/b}=q_{0j}^{a/b}$ and $p_{1j}^{a/b}=q_{1j}^{a/b}, $ then $\delta _j(a, b)=0.$ 
This is the conventional probabilistic rule that we have in classical physics
and in quantum physics in the absence of interference. 

If perturbations produced by preparations of ensembles $\tilde{\Omega}_i^b$ (via filtrations of 
$\Omega$)\footnote{In general we need two copies of $\Omega$ to prepare $\tilde{\Omega}_0^b$ and $\tilde{\Omega}_1^b.$}
are such that $\delta_j\neq 0,$ we obtain `nonconventional probabilistic rules'.  

The coefficients $\delta_i$ and $\lambda_i$  are called {\it statistical deviations}
and {\it normalized statistical deviations,} respectively. The magnitudes of normalized statistical deviations
will play the cruicial role in our further considerations.

First we remark that if (for $j=0,1)$
\begin{equation}
\label{M}
|\lambda_j|\leq 1
\end{equation}
and the matrix of probabilities  $(q_{ij}^{a/b})$ is double stochastic,
we obtain the quantum probabilistic transformation (\ref{q0}), (\ref{q})
by choosing  $\lambda_0=\cos \theta$ and $\lambda_1=-\cos \theta.$ 

We now present the general classification
of probabilistic rules in nature. The general transformation of probabilities
for three ensembles $\Omega, \tilde{\Omega}_i^b, i=0,1,$ has the form:
\begin{equation}
\label{MA}
p_j^a=p_0^b q_{0j}^{a/b} + p_1^b q_{1j}^{a/b} + 2\sqrt{p_0^b p_1^b q_{0j}^{a/b}q_{1j}^{a/b}} 
\lambda_j, 
\end{equation}
where $\lambda_j, j=0,1,$ are given by (\ref{L}).

Depending on the magnitudes of normalized statistical deviations, we can obtain:

1) the trigonometric probabilistic rule, $\lambda_j=\cos 
\theta_j, j=0,1;$ 

2) hyperbolic probabilistic rule, 
$\lambda_j=\pm \cosh \theta_j, j=0,1;$

3) hyper-trigonometric probabilistic rule,
for example, 

$\lambda_0 = \pm \cosh \theta_1, 
\lambda_1 = \cos \theta_2.$ 

Here `phases' $\theta$ are just special parameterizations for normalized statistical deviations.

In particular, the trigonometric probabilistic rule contains the conventional 
probabilistic rule, $\lambda_j=0,$ and the quantum probabilistic rule (under the additional condition that 
($q^{a/b}_{ij}$) is a double stochastic matrix). Here normalized statistical deviations are relatively small,
see (\ref{M}). In the hyperbolic case they are quite large, namely:
\begin{equation}
\label{M1}
|\lambda_j|\geq 1
\end{equation}
for both $j=0,1.$ Here the order of perturbations via filtrations is 
essentially larger than, in particular,  in quantum experiments.

\medskip

{\bf Example 1.}  Let ${\cal E},{\cal E}_0,{\cal E}_1$ 
produce following perturbations of probabilities: 
$$
\Delta_{0j}\equiv(p_{0j}^{a/b} - q_{0j}^{a/b}) p_0^b 
=2\xi_{0j}\sqrt{p_0^b p_1^b 
q_{0j}^{a/b} q_{1j}^{a/b}}; 
$$ 
$$
\Delta_{1j} = (p_{1j}^{a/b}-q_{1j}^{a/b})p_1^b=2\xi_{1j}\sqrt{p_0^b p_1^b 
q_{0j}^{a/b} q_{1j}^{a/b}}.
$$
where $\xi_{0j}$ and $\xi_{1j}$ are some coefficients. These 
coefficients determine the corresponding transformation rule for probabilities. If $\xi_{0j}
+\xi_{1j}=0,$ we obtain the conventional  rule. If $\xi_{0j}+\xi_{1j}\neq 0,$ 
we obtain nonconventional rules; in 
particular, in quantum theory we have $|\xi_{0j}+\xi_{1j}|\leq 1.$

\medskip

In a mathematical model we can describe filters $F_i, i=0, 1,$ by two maps 
$$
g_0:\Omega_0^b \rightarrow \Omega_0^b, g_1: \Omega_1^b \rightarrow \Omega_1^b
$$

The pair of maps $g_0$ and $g_1$ induces the map 

$g: \Omega \rightarrow \Omega.$ 

In principle, we can 
consider a more general model in that filters $F_i$ can also change property $b$ (some systems with 
$b=0$ can be transformed into systems with $b=1$ and vice versa).
In such a model we have to consider general 
transformations $g:\Omega \rightarrow \Omega$ 
(i.e., without the restriction $g(\Omega_i^b)=\Omega_i^b$). We plan to study this model in further papers.

We shall use the symbols $\tilde{\Omega}_0^b$ and $\tilde{\Omega}_1^b$ to 
denote ensembles $g(\Omega_0^b)$ and $g(\Omega_1^b).$ We remark that in our
model (with $g(\Omega_j^b)=\Omega_i^b$) new ensembles coincide with original 
ensembles as collections of physical systems (for example, particles), but physical properties are changed by filtrations.

Probability ${\bf P}_{\Omega^b_i}$ on the ensemble $\Omega^b_i$ (conditional probability
${\bf P}_{\Omega} (\cdot/ \Omega^b_i))$ is lifted to ensemble $\tilde{\Omega}_i^b$ with the aid of map 
$g_i:$
$$
{\bf P}_{\tilde{\Omega}^b_i}(O) = {\bf P}_{\Omega^b_i}(g_i^{-1}(O)).
$$
We remark that by definition of conditional probability this probability is equal to
${\bf P}_\Omega(g_i^{-1}(O) \bigcap \Omega_i^b)/{\bf P}_\Omega(\Omega_i^b).$ Moreover, as $g_i$ maps
$\Omega_i^b$ on itself, we get that this probability is equal to :

${\bf P}_\Omega(g_i^{-1}(O))/{\bf P}_\Omega(\Omega_i^b).$ 

We note that the 
probability ${\bf P}_{\tilde{\Omega}_i^b}$ is nothing than the $g_i$-image of the conditional 
probability ${\bf P}_{\Omega_i^b}:\;
{\bf P}_{\tilde{\Omega}_i^b} = g_i^*{\bf P}_{\Omega_i^b}.$

It is assumed that ensembles $\tilde{\Omega}_i^b$ are equipped by 
some $\sigma$-algebras and maps $g_i$ are measurable.

We note that in  quantum theory
probability distributions  ${\bf P}_{\varphi_i}$ (for states $\varphi_i)$
are nothing than ${\bf P}_{\tilde{\Omega}_i^b}, i=0,1.$
\footnote{Later we shall present concrete maps $g_i$ that reproduce quantum probability rule.}

In general we introduce probabilities 
$$
q_{ij}^{a/b} \equiv {\bf P}_{\tilde{\Omega}_i^b}(a=j)=
{\bf P}_{\Omega_i^b} (a(g_i(\omega))=j).
$$
By definition of conditional probability
$$
q_{ij}^{a/b} ={\bf P}_\Omega(a(g_i)(\omega))=j/b(\omega)=i)
=\frac{q_{ij}^{ab}}{p_i^b}, i, j=0, 1,
$$
where

$q_{ij}^{ab}={\bf P}_\Omega(a(g_i)(\omega))=j, b(\omega)=i).$

In particular, in quantum theory $q_{ij}^{a/b}={\bf P}_{\varphi_i}(a=j).$

The coefficients $\lambda_j(a, b)$ corresponding to perturbations 
of probabilities by filtrations can be represented as
\begin{equation}
\label{LA}
\lambda_j(a,b)=\frac{(p_{0j}^{ab}-q_{0j}^{ab})+ (p_{1j}^{ab}-q_{1j}^{ab})}{2{\sqrt{q_{0j}^{ab} q_{1j}^{ab}}}}
\end{equation}

{\bf Remark.} (Contextualism) In the contextualist
HV-model preparation procedures ${\cal E}_0$ and ${\cal E}_1$
disturb not physical variable $a,$ but hidden variable $\omega.$

{\bf Remark.} (Contextualism without HV) {\small We can even forget about HV and consider 
$\omega$ as purely mathematical `chance parameter.' In this framework
all our considerations are based on the observation that measurements of 
the $a$-variable for three different statistical ensembles $\Omega,$
$\tilde{\Omega}_0^b$ and $\tilde{\Omega}_1^b$ are performed via three
different measurement procedures. Therefore they are described by 
different random variables $a(\omega), \tilde{a}^{(0)}(\omega)= a(g_0(\omega)),
\tilde{a}^{(1)}(\omega)= a(g_1(\omega)).$  We ask ourself: What kind of transformations
connecting probability distributions of these random variables can we obtain in different
experiments? It seems that it is too general formulation of the problem (however, see remark
at the end of the paper). It would be more natural to consider only transformations
that are perturbations of the standard formula of total probability (based on the Bayes' rule
for conditional probabilities). In fact, we proved that there are only three types of such 
probabilistic transformations: trigonometric, hyperbolic and hyper-trigonometric. The creation of 
quantum formalism was the important discovery in probability theory: it was found (on the basis
of interference experiments) that in some experimental situations the formula of total probability
has to be disturbed by trigonometric perturbation. However, the probabilistic roots of this perturbation
term were not found. This implied the creation of the model of micro-reality based on 
the wave-particle duality. The possibility of non-trigonometric interferences was not observed.}

\section{Simulation of trigonometric and hyperbolic interference}

Measurements producing trigonometric and hyperbolic interferences that are presented
in this section can be easily simulated by using just a pseudo-random generator of numbers
uniformly distributed on the segment $[0,1].$

Let $\Omega=[0,1], {\cal F}$ - the $\sigma$-algebra of Borel sets, ${{\bf P}}=dx$ is the uniform probability distribution (Lebesque measure). Let

\[
c(\omega)=\left\{ \begin{array}{ll}
1, x\in[0,\frac{1}{3})\\
0, x\in [\frac{1}{3}, 1]
\end{array}
\right.
\]

\[
a(\omega)=\left\{ \begin{array}{ll}
1, x\in(\frac{1}{4}, \frac{3}{4}]\\
0, x\not\in (\frac{1}{4}, \frac{3}{4}]
\end{array}
\right.
\]

So we have ensembles $\Omega_1^b=[0,\frac{1}{3}), \Omega_0^b=[\frac{1}{3}, 1];$
and probabilities:

$p_1^b={\bf P}_\Omega(\Omega_1^b)=\frac{1}{3}, \; p_0^b={{\bf P}}_\Omega(\Omega_0^b)=\frac{2}{3}; $
$$
p_{01}^{ab}={{\bf P}}_\Omega(a(\omega)=1, b(\omega)=0)=\frac{5}{12}, 
p_{00}^{ab}={{\bf P}}_\Omega(a(\omega)=0, b(\omega)=0)=\frac{1}{4}, 
$$
$$
p_{11}^{ab}={{\bf P}}_\Omega(a(\omega)=1, b(\omega)=1)=\frac{1}{12}, 
p_{10}^{ab}={{\bf P}}_\Omega(a(\omega)=0, c(\omega)=1)=\frac{1}{4}.
$$

Let us consider maps $g_i$ which describe changes of $a$ in the processes of 
filtrations: 

$g_0:\Omega_0^b \rightarrow \Omega_0^b$ such that  $g_0([\frac{1}{3}, \beta ])=[\frac{1}{3}, 
\frac{3}{4}]$ and $g_0((\beta , 1])=(\frac{3}{4}, 1].$ 

So $\tilde{\Omega}_0^b=g_0(\Omega_0^b)=
[\frac{1}{3}, 1].$

$g_1:\Omega_1^b \rightarrow \Omega_1^b$ such that $g_1([0, \alpha])=[0, \frac{1}{4}]$ and 
$g_1((\alpha, \frac{1}{3}))=(\frac{1}{4}, \frac{1}{3}).$ 

So $\tilde{\Omega}_1^b=g_1(\Omega_1^b)=[0, \frac{1}{3})$.

Here $\alpha$ and $\beta$ are parameters such that $0<\alpha<\frac{1}{3}$ and $\frac{1}{3}<\beta<1.$

Of course, there exist various maps $g_0$ and $g_1$ which satisfy the above 
conditions. Different maps correspond to different physical realizations of filters. We shall see 
that by varying parameters $\alpha$ and $\beta$ we can obtain classical, trigonometric (in 
particular, quantum) and hyperbolic probabilistic behaviours. We have 

$q_{01}^{ab}={{\bf P}}_\Omega(a(g_0(\omega))=1, b(\omega)=0)=\beta-\frac{1}{3};$ 

in the same 
way $q_{00}^{ab}=1-\beta; q_{11}^{ab}=\frac{1}{3}-\alpha; q_{10}^{ab}=
\alpha.$

To generate quantum behaviour, we have to have a double stochastic matrix of probabilities 
($q_{ij}^{a/b}$).

We have: $q_{ij}^{a/b}={{\bf P}}_{\tilde{\Omega}_i^b}(a=j).$ So 

$q_{01}^{a/b}=
\frac{3\beta-1}{2}; q_{00}^{a/b}=\frac{3-3\beta}{2}; q_{11}^{a/b}=1-3\alpha,  q_{10}^{a/b}=3\alpha.$ 

The matrix $(q_{ij}^{a/b})$ is always stochastic: 

$q_{00}^{a/b}+q
_{01}^{a/b}={{\bf P}}_{\tilde{\Omega}_0^b}(a=0)+ {{\bf P}}_{\tilde{\Omega}_0^b}(a=1)=1(=\frac{3-3\beta}{2}
+\frac{3\beta-1}{2});$ 

$q_{10}^{a/b}+q_{11}^{a/b}={{\bf P}}_{\tilde{\Omega}_1^b}(a=0)+{{\bf P}}
_{\tilde{\Omega}_1^b}(a=1)=1(=1-3\alpha+3\alpha).$ 

It is double stochastic if 

$q_{00}^{a/b}+q_{10}^{a/b}=\frac{3-3\beta}{2}+3\alpha=1; q_{01}^{a/b}+q_{11}^{a/b}=\frac{3\beta-1}
{2}+1-3\alpha=1.$

Thus
\begin{equation}
\label{DS}
\beta=\frac{1}{3}+2\alpha
\end{equation}

(we note that if $\alpha$ varies between 0 and 1/3, then $\beta$ varies between 1/3 and 1).

Double stochasticity implies:
$q_{01}^{a/b}=q_{10}^{a/b}=3\alpha$ and $q_{00}^{a/b}=q_{11}^{a/b}=1-3\alpha.$ 
In general by using formula (\ref{LA}) we get 
$$\lambda_0=\frac{(\beta-\frac{3}{4})+(\frac{1}{4}-\alpha)}{2\sqrt{\alpha(1-\beta)}}=
\frac{(\beta-\alpha-\frac{1}{2})}{2\sqrt{\alpha(1-\beta)}}; 
$$
$$\lambda_1=\frac{(\frac{1}{2}+\alpha-\beta)}{2\sqrt{(\beta-\frac{1}{3})(\frac{1}{3}-\alpha)}}.
$$
If $\beta=\alpha+\frac{1}{2},$ we generate conventual probabilistic rule in that $\lambda_0=\lambda_1=0.$

In particular,  we get this rule in the case of filtrations such that: $\alpha=1/4$ and $\beta=3/4.$  
In particular, such filtrations can be realized by identical maps: $g_0(x)=x, g_1(x)=x.$ These filters
are ideal: they do not perturb the $a$-variable at all. However, there are infinitely many
filters $g_0, g_1$ with $\alpha=1/4$ and $\beta=3/4$  that produce essential perturbations for individual
physical systems: the variable $a$ can be strongly  changed in some points. Nevertheless, these filters
do not produce interference of probabilistic alternatives, since statistical deviations are negligibly 
small (compare with Ballentine's analysis of Heisenberg uncertainty principle, [15], see also [13]).

We want to obtain nontrivial quantum behaviour: 
$\lambda_0=\cos\theta_0=-\lambda_1\neq0. $ First we use the 
condition of double stochasticity (\ref{DS}): 
$$\lambda_0=\frac{\alpha-\frac{1}{6}}{2{\sqrt{\alpha(\frac{2}{3}-2\alpha)}}}=-\lambda_1.
$$

We obtain trigonometric interference if
$$
|\lambda_0|=\frac{|\alpha-\frac{1}{6}|}{2\sqrt{2\alpha(\frac{1}{3}-\alpha)}} \leq 1.
$$

Direct computation demonstrates that $\lambda_0^{\prime}(\alpha)>0$ for $\alpha\in(0,1/3).$
Thus the interference term $\lambda_0(\alpha)$ is the 
increasing function of the perturbation parameter 
$\alpha$ and $\lambda_0(0)=-\infty, \lambda_0(\frac{1}{3})=+\infty, \lambda_0(\frac{1}{6})=0, 
\lambda_0(\alpha_\pm)=\pm1,$ where $\alpha_\pm=\frac{3\pm2\sqrt{2}}{18}.$

Therefore perturbations (due to filtration) varying in the segment 
$[\alpha_-, \alpha_+]$ 
generate quantum behaviour: $\lambda_0(\alpha)=\cos \theta_0$ is varying from $-1$ to 1. 
The phase-angle $\theta_0$ can vary from $[\pi, 2\pi]$ (by a parameterization 
of the perturbation variable we can get $\theta_0$ varying from 0 to $\pi$). 

{\bf Remark.} (Contextualist model with hidden variable) {\small We can use the contextualist
interpretation of quantum mechanics.
Here $\alpha$ and $\beta$ are parameters determining the context of an experiment, $\omega$ is 
a hidden variable.}

We now consider the extreme values of $\alpha:\alpha=\alpha_\pm$
(maximal interference) and $\alpha=\frac{1}{6}$ (decoherence).
Quantities 
$\pi_0=\frac{1}{4}-\alpha$ and $\pi_1=\frac{3}{4}-\beta$ 
can be used as measures of perturbation of $a$ due to the transitions
\begin{equation}
\label{T}
\Omega_0^b\rightarrow\tilde{\Omega}_0^b \; \; {\rm{ and }}  \; \; \Omega_1^b\rightarrow\tilde{\Omega}_1^b.
\end{equation}

Let $\alpha=1/6.$ Here $\pi_0=\frac{1}{12}; \pi_1=\frac{1}{12}.$ The absence of interference is characterized by the symmetric shift of the $a$-property under the transition (\ref{T}).

We remark that there exists only one point $\alpha$ that satisfies to the decoherence condition:
$\pi_0=\pi_1,$ namely $\alpha=1/6$ (under condition $\beta=\frac{1}{3}+2\alpha$). 

We have to differ `classical situation' from `quantum decoherence'. 
In the first case $\delta(a, b)=0$ as the result of the precise 
filtration: $\alpha=\frac{1}{4}$ and $\beta=\frac{3}{4}.$ In the second case filtration
induces essential perturbations. However, these perturbations compensate each other. 

In fact, the condition of double stochasticity (see(\ref{DS})) plays here the cruicial 
role: it is a kind of constraint between perturbations induced by preparations of ensembles
$\tilde{\Omega}_0^b$ and $\tilde{\Omega}_1^b.$ It seems that this condition is the root of `quantum mystery'. On the other hand, $\cos \theta -$factor (interference) can be induced by `classical preparation procedures' for macro-systems.

The quantity 

$\xi(\alpha)=|\pi_0(\alpha)-\pi_1(\alpha)|=|\alpha-\frac{1}{6}|$

can be used as
a measure of asymmetry of perturbations (\ref{T}). In the case of decoherence 
$\xi=\xi(\frac{1}{6})=0.$ For $\alpha\in[\alpha_-, \alpha_+],
\xi(\alpha)=|\alpha-1/6|$ yields the maximal value for $\alpha=\alpha_\pm.$ Here 
$\xi=\xi_{\max}=\frac{\sqrt{2}}{9}.$ Thus maximum of interference ($|\cos\theta_0|=1|$) corresponds to maximal asymmetry of perturbations.

What kind of behaviour would we observe for perturbations $\alpha\not\in(\alpha_-, \alpha_+)$?
We see that, instead of `trigonometric quantum behaviour', we obtain `hyperbolic quantum behaviour'.

If $\alpha>\alpha_+$ (but $\alpha<1/3),$ then $\lambda_0(\alpha)$ can be represented as 
$\cosh\theta_0;$ if the perturbation parameter $\alpha$ is varying from $\alpha_+$to 1/3, 
then $\cosh\theta_0$ is varying from 1 to + $\infty.$ So the `
hyperbolic phase' can be chosen belonging to $[0, +\infty).$ If 
$\alpha<\alpha_-$ (but $\alpha>0), $ then $\lambda_0(\alpha)$ can be represented as $-\cosh\theta_0.$ 
The variation of $\alpha$ from  $\alpha_-$ to $0,$ 
implies the variation $\lambda_0$ from -1 to  $- \infty.$ So $\theta_0\in[0, \infty).$

`Hyperbolic quantum behaviour' may be induced by filters having sufficiently 
strong perturbations. In our model there are the thresholds $\alpha=\alpha_{\pm}.$ At these levels of perturbations `quantum trigonometric behaviour' is transformed to `quantum hyperbolic behaviour'. Our model gives the possibility to simulate such 
a transition by using macro-systems. It may be that such a transition can be observed for some `natural' physical processes.

The formalism of hyperbolic quantum mechanics (representation of the probabilistic rule (\ref{QA}) in a linear
space) was developed in [34]. Instead of complex numbers, we have to work with so called hyperbolic numbers, 
see [35], p. 21. The development of hyperbolic quantum mechanics can be interesting for
comparative analysis with standard quantum mechanics. In particular, we clarify
the role of complex numbers in quantum theory. Complex (as well as hyperbolic)
numbers were used to linearize nonlinear probabilistic rule (that in general
could not be linearized over real numbers). Another interesting feature of
hyperbolic quantum mechanics is the violation of the principle of superposition.
Here we have only some restricted variant of this principle. 

We remark that the quantum probabilistic transformation

${\bf P}= {\bf P}_1 + {\bf P}_2 + 2 \sqrt{{\bf P}_1 {\bf P}_2} \cos \theta$

gives the possibility to predict the probability ${\bf P}$ if we
know probabilities ${\bf P}_1$ and ${\bf P}_2.$ In principle,
there might be created theories based on arbitrary transformations:
$$
{\bf P}= F({\bf P}_1, {\bf P}_2).
$$
It may be that some rules have linear space representations
over `exotic number systems', for example, $p$-adic numbers [36].

\medskip

Preliminary analysis of probabilistic foundations of quantum mechanics
(that induced the present investigation) was performed in the books [18]
and [36] (chapter 2);
a part of results of this paper was presented in preprints [37],[38].

\section*{Acknowledgements}
I would like to thank 
S. Albeverio, L. Accardi, L. Ballentine, V. Belavkin, E. Beltrametti, W. De Muynck,
S. Gudder, T. Hida, A. Holevo,  P. Lahti,  A. Peres, J. Summhammer,  
I. Volovich for (sometimes critical) discussions on probabilistic foundations of quantum 
mechanics.

\bigskip

{\bf References}

[1] P. A. M.  Dirac, {\it The Principles of Quantum Mechanics}
(Claredon Press, Oxford, 1995).

[2] N. Bohr, {\it Phys. Rev.,} {\bf 48}, 696 (1935).

[3] J. von Neumann, {\it Mathematical foundations
of quantum mechanics} (Princeton Univ. Press, Princeton, N.J., 1955).

[4] E. Schr\"odinger, {\it Die Naturwiss}, {\bf 23}, 807-812, 824-828, 844-849 (1935).

[5] E. Schr\"odinger,  {\it What is an elementary particle?} in Gesammelte Abhandlungen.
(Wieweg and Son, Wien 1984). 

[6] W. Heisenberg, {\it Z. Physik.,} {\bf 43}, 172 (1927).

[7] R. Feynman and A. Hibbs, {\it Quantum Mechanics and Path Integrals}
(McGraw-Hill, New-York, 1965).

[8] E. Schr\"odinger, {\it Philosophy and the Birth of Quantum Mechanics.}
Edited by M. Bitbol, O. Darrigol (Editions Frontieres, 1992).

[9] B. d'Espagnat, {\em Veiled Reality. An anlysis of present-day
quantum mechanical concepts} (Addison-Wesley, 1995). 

[10] A. Peres, {\em Quantum Theory: Concepts and Methods} (Kluwer Academic Publishers, 1994).

[11] J. M. Jauch, {\it Foundations of Quantum Mechanics} (Addison-Wesley, Reading, Mass., 1968).

[12]  P. Busch, M. Grabowski, P. Lahti, {\it Operational Quantum Physics}
(Springer Verlag, 1995).

[13]   W. De Muynck, W. De Baere, H. Martens,
 {\em Found.  Phys.} {\bf 24}, 1589--1663 (1994).

[14]  L. E. Ballentine, {\it Quantum mechanics} (Englewood Cliffs, 
New Jersey, 1989).

[15]  L. E. Ballentine, 
{\it Rev. Mod. Phys.}, {\bf 42}, 358--381 (1970).

[16] L. Accardi, The probabilistic roots of the quantum mechanical paradoxes.
{\em The wave--particle dualism.  A tribute to Louis de Broglie on his 90th 
Birthday,} ed. S. Diner, D. Fargue, G. Lochak and F. Selleri
(D. Reidel Publ. Company, Dordrecht, 297--330, 1984).

[17] J. Summhammer, {\em Int. J. Theor. Phys.} {\bf 33}, 171-178 (1994).

[18] A.Yu. Khrennikov, {\it Interpretations of 
probability} (VSP Int. Publ., Utrecht, 1999).

[19] A. Einstein, B. Podolsky, N. Rosen,  Phys. Rev., {\bf 47}, 777--780
(1935).

[20] J.S. Bell,
   Rev. Mod. Phys., {\bf 38}, 447--452 (1966).
 
[21] J.F. Clauser , M.A. Horne, A. Shimony, R. A. Holt,
 Phys. Rev. Letters, {\bf 49}, 1804-1806 (1969);
 J. S. Bell, {\it Speakable and unspeakable in quantum mechanics.}
 (Cambridge Univ. Press, 1987).
 J.F. Clauser ,  A. Shimony,  Rep. Progr.Phys.,
 {\bf 41} 1881-1901 (1978).
  A. Aspect,  J. Dalibard,  G. Roger, 
 Phys. Rev. Lett., {\bf 49}, 1804-1807 (1982);
  D. Home,  F. Selleri, Nuovo Cim. Rivista, {\bf 14},
 2-176 (1991).
 
 [22] A. Shimony, {\it Search for a naturalistic world view.} (Cambridge Univ. Press, 1993).

 [23] H. P. Stapp, Phys. Rev., D, {\bf 3}, 1303-1320 (1971);
 P.H. Eberhard, Il Nuovo Cimento, B, {\bf 38}, N.1, 75-80(1977); Phys. Rev. Letters,
 {\bf 49}, 1474-1477 (1982);
 A. Peres,  Am. J. of Physics, {\bf 46}, 745-750 (1978).
 P. H. Eberhard,  Il Nuovo Cimento, B,
 {\bf 46}, N.2, 392-419 (1978).

[24] I. Pitowsky,  Phys. Rev. Lett, {\bf 48}, N.10, 1299-1302 (1982);
  Phys. Rev. D, {\bf 27}, N.10, 2316-2326 (1983);
  S. P. Gudder,  J. Math Phys., {\bf 25}, 2397- 2401 (1984);
  S. P. Gudder, N. Zanghi,
  Nuovo Cimento B {\bf 79}, 291--301 (1984).

 [25]  A. Fine,  Phys. Rev. Letters, {\bf 48}, 291--295 (1982);
 P. Rastal, Found. Phys., {\bf 13}, 555 (1983).
 W. Muckenheim,  Phys. Reports, {\bf 133}, 338--401 (1986);
 W. De Baere,  Lett. Nuovo Cimento, {\bf 39}, 234-238 (1984);
 {\bf 25}, 2397- 2401 (1984); W. De Muynck and W. De Baere W.,
  Ann. Israel Phys. Soc., {\bf 12}, 1-22 (1996); 
  W. De Muynck, J.T. Stekelenborg,  Annalen der Physik, {\bf 45},
 N.7, 222-234 (1988).

[26] A. Yu. Khrennikov, {\it J. of Math. Physics,} {\bf 41}, N.4, 1768-1777 (2000);
 {\it Il Nuovo Cimento,} {\bf B 115}, N.2, 179-184 (2000); {\it J. of Math. Physics}, {\bf 41}, N.9, 5934-5944 (2000).

[27] L. De Broglie, {\it The current interpretation of wave mechanics. 
A critical study.} (Elsevier Publ. co., Amsterdam, 1964).

[28]	D. Bohm  and B. Hiley, {\it The undivided universe:
an ontological interpretation of quantum mechanics.} 
(Routledge and Kegan Paul, London, 1993).

[29] S. D'Agostino, Continuity and completeness in physical theory:
Schr\"odinger's return to the wave interpretation of 
quantum mechanics in the 1950's; 339- 360 in 
{\it E. Schr\"odinger,  Philosophy and the Birth of Quantum Mechanics.}
Edited by M. Bitbol, O. Darrigol (Editions Frontieres, 1992).

[30] M. Lockwood, What Schr\"odinger should have learned from his cat;
363- 384 in 
{\it E. Schr\"odinger,  Philosophy and the Birth of Quantum Mechanics.}
Edited by M. Bitbol, O. Darrigol (Editions Frontieres, 1992).

[31] A. Fine, {\it The shaky game.} (University of Chicago press, Chcago/London, 1988).

[32] W. M. De Muynck, Interpretations of quantum mechanics,
and interpretations of violations of Bell's inequality. To be published
in {\it Proceedings of the Conference "Foundations of Probability and Physics",}
V\"axj\"o, Sweden-2000.

[33] A. N. Kolmogoroff, {\it Grundbegriffe der Wahrscheinlichkeitsrechnung}
(Springer Verlag, Berlin, 1933); reprinted:
{\it Foundations of the Probability Theory}. 
(Chelsea Publ. Comp., New York, 1956).

[34] A. Yu. Khrennikov, {\it Hyperbolic quantum mechanics.} Preprint quant-ph/0101002, 31 Dec (2000). 

[35] A. Yu. Khrennikov, {\it Supernalysis}  (Kluwer Academic Publishers, 
Dordreht, 1999).

[36]  A. Yu. Khrennikov, {\it $p$-adic valued distributions in 
mathematical physics} (Kluwer Academic Publishers, Dordrecht, 1994).

[37] A. Yu. Khrennikov, {\it Ensemble fluctuations and the origin of quantum probabilistic
rule.} Rep. MSI, V\"axj\"o Univ., {\bf 90}, October (2000).

[38] A. Yu. Khrennikov, {\it Classification of transformations of  probabilities for preparation procedures:
trigonometric and hyperbolic behaviours.} Preprint quant-ph/0012141, 24 Dec (2000).

\end{document}